\documentclass[aps,pre,showpacs,amsmath,amssymb,amsfonts,superscriptaddress,twocolumn]{revtex4-1}

\usepackage{graphicx}
\usepackage{subfigure}
\usepackage{verbatim}
\usepackage{dcolumn}
\usepackage{bm}
\usepackage{epsf}
\usepackage{color}
\usepackage[colorlinks=true,citecolor=blue,linkcolor=blue,urlcolor=blue]{hyperref}
\usepackage{lmodern}
\usepackage{tikz}
 
\newcommand{\bra}[1]{\left\langle #1\right|}
\newcommand{\ket}[1]{\left|#1\right\rangle}

\newcommand{\tr}[1]{\mathrm{tr}\left\{#1\right\}}

\newcommand{\la}{\left\langle}
\newcommand{\ra}{\right\rangle}

\newcommand{\td}{\mathrm{d}}

\newcommand{\e}[1]{\exp{\left(#1\right)}}

\newcommand{\bla}{bla\\bla\\bla\\bla\\bla}

\newcommand{\mc}[1]{\mathcal{#1}}

\newcommand{\mrm}[1]{\mathrm{#1}}

\newcommand{\w}{\omega}
\newcommand{\s}{\sigma}
\newcommand{\dg}{\dagger}
\newcommand{\sig}[1]{\sigma_{#1}}

\begin{document}

\title{Thermodynamic universality of quantum Carnot engines}

\author{Bart\l{}omiej Gardas}
\affiliation{Theoretical Division, Los Alamos National Laboratory, Los Alamos, NM 87545, USA}
\affiliation{Institute of Physics, University of Silesia, 40-007 Katowice, Poland}

\author{Sebastian Deffner}
\email{sdeffner@lanl.gov}
\affiliation{Theoretical Division, Los Alamos National Laboratory, Los Alamos, NM 87545, USA}
\affiliation{Center for Nonlinear Studies, Los Alamos National Laboratory, Los Alamos, NM 87545, USA}

\date{\today}

\begin{abstract}
The Carnot statement of the second law of thermodynamics poses an upper limit on the efficiency of all heat engines. Recently, it has been studied whether generic quantum features such as coherence and quantum entanglement could allow for quantum devices with efficiencies larger than the Carnot efficiency. The present study shows that this is not permitted by the laws of thermodynamics -- independent of the model. We will show that rather the definition of heat has to be modified to account for the thermodynamic cost for maintaining non-Gibbsian equilibrium states. Our theoretical findings are illustrated for two experimentally relevant examples.
\end{abstract}


\maketitle

\section{Introduction}

Harnessing energy stored in inaccessible forms such as heat or chemical energy and transforming it into useful work is one of the most important, technological achievements. Nevertheless, the underlying principles and ultimate limitations imposed by quantum mechanics on such thermodynamic processes are still an active field of research \cite{Gemmer2009a,Kosloff2013c}. Modern studies range from implementations of quantum heat engines in ion traps \cite{abah_2012} over thermodynamic cycles in optomechanical 
systems~\cite{Zhang_2014,Gelbwaser-Klimovsky2015} to the description of the principles of photosynthesis as photo-Carnot engines~\cite{Castro_2014,Romero_2014}. The natural question arises how generic quantum features such as coherence and entanglement affect classical formulations of the thermodynamic axioms.
This problem has  been studied from many different perspectives~\cite{Goold} including stochastic thermodynamics~\cite{Parrondo,Verley2014a,Proesmans2015} and information theory~\cite{Terry,Rudolph,Oppenheim_2013}. However, a conclusive consensus appears still to be lacking.

In particular, it has been studied whether quantum correlations could be harnessed, and whether quantum devices could operate with efficiencies larger than the Carnot efficiency~\cite{scully_2003,Lutz_2009, HTQ, Paolo} -- therefore demanding a reformulation of
the Carnot statement of the second law of thermodynamics.

The Carnot statements of the second law of thermodynamics declares~\cite{carnot_24}
\begin{center}
\textit{No engine operating between two heat reservoirs can be more efficient than a Carnot engine operating between those same reservoirs.}
\end{center}
Recent studies, however, raised the question whether quantum effects such as coherence and entanglement could provide means to break the limit posed by the Carnot efficiency~\cite{scully_2003,Lutz_2009,Kurizki_2013,Oppenheim_2013}. To this end, a variety of theoretical and experimental setups have been developed~\cite{abah_2012,Zhang_2014,Amikam_2014}. 

Thermal equilibrium states of classical systems are universally described by the Boltzmann-Gibbs distribution~\cite{callen,Jin_2013}. It can be easily shown that the Carnot statement is only a consequence of this universality of equilibrium states~\cite{callen}. The situation is dramatically different for thermally open quantum systems. Generically, quantum systems, which are not only ultra-weakly coupled to their environment, do not relax into Gibbs states \cite{Gelin2009,Barkai_2014}. This can be seen most clearly for the analytically solvable case of quantum Brownian motion \cite{Hu1992}.  It has been shown that the  quantum correlations between system and environment prevent relaxation into the Gibbs equilibrium states \cite{Horhammer2008}.  Therefore, such non-Gibbsian equilibrium states are not fully ``thermalized'' and contain additional information, encoded in quantum coherence and entanglement~\cite{hor_2009,vedral_2012}.
\begin{figure}
\includegraphics[width=.48\textwidth]{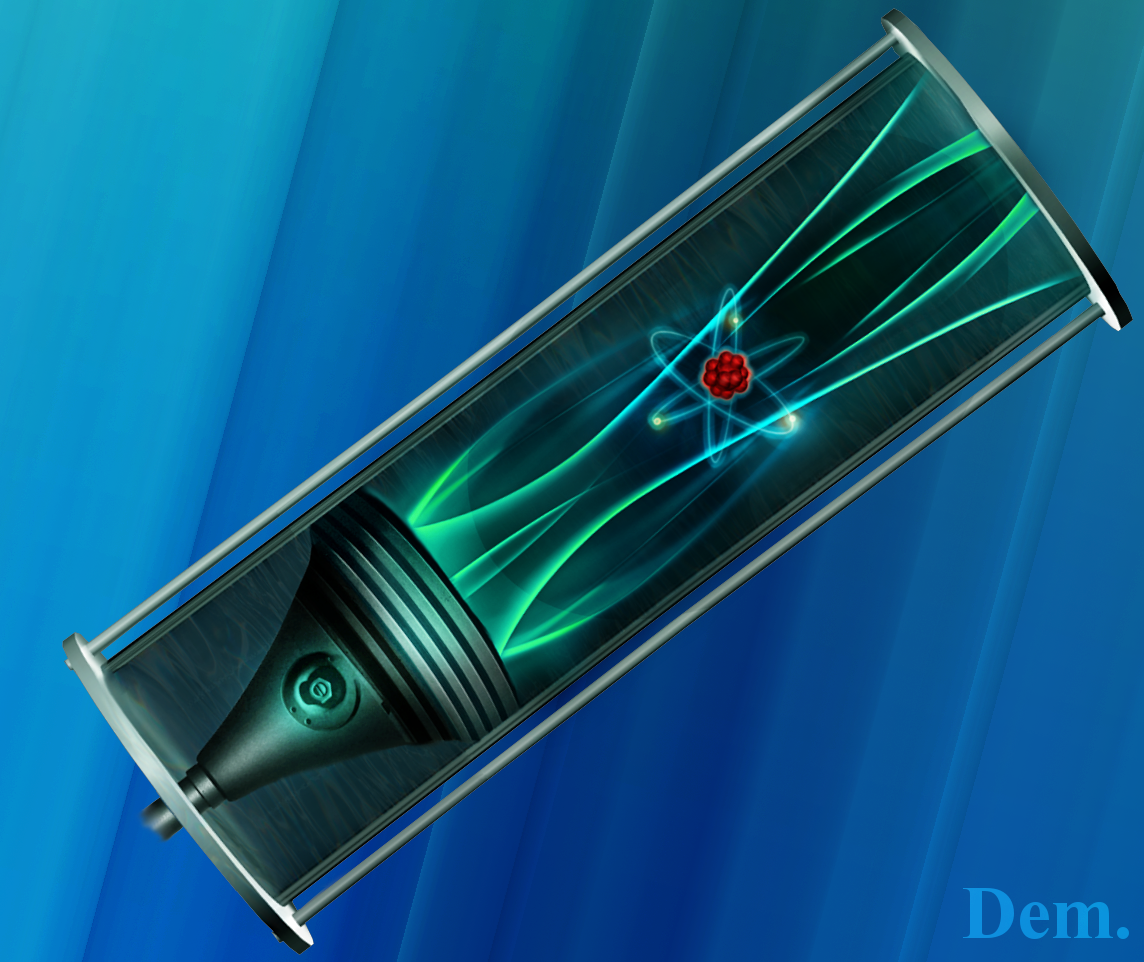}
\caption{\label{fig:model} Illustration of a generic quantum heat engine: A quantum particle inside in a quantum piston consisting, e.g., of an optical cavity.}
\end{figure}

It is only natural to ask whether a quantum heat engine such as in Fig.~\ref{fig:model} could be realized that utilizes this ``extra" information and thus constitutes a device operating with an efficiency larger than predicted by Carnot \cite{scully_2003}. The present study will elucidate that this is not permitted by the laws of thermodynamics. In particular, we will argue that there is a specific thermodynamic price that has to be to paid to maintain quantum correlations and thus to prevent the system from relaxing into a Gibbs state. How to properly modify the definition of heat has been recently studied with great intensity for various model systems \cite{Kurizki,Gelbwaser-Klimovsky2015,Yamamoto2015,Zheng2015,Lin2015}. However, to the best of our knowledge, the only rigorous and model independent distinction has  been previously discussed  in the context of the heat capacity \cite{Hanggi2008} of open quantum systems (see Eq. (23) of Ref.~\cite{Hanggi2008}).

In the present work we  further develop this notion of quantum heat. As a main result we will see that the classical Carnot statement remains unchanged for open quantum systems with arbitrary coupling to their environment. More specifically, we will show
\begin{center}
\textit{No quantum heat engine operating in quasistatic Carnot cycles can harness quantum correlations.}
\end{center}
This insight has far-reaching consequences for all areas of engineering at the nanoscale. Nanoengines performing beyond the Carnot limit necessarily operate  far from thermal equilibrium \cite{Lutz_2014}. To the best of our knowledge, however, this has only been proven rigorously, if at least the initial state is Gibbsian~\cite{Alicki1979,sagawa,Campisi2014}.

\section{Quasistatic processes}

Before we analyze the quantum Carnot cycle, let us establish an important concept, first. Consider a quantum system with  Hamiltonian $H(\w_t)$, where $\w_t$ is an external control parameter. Then the dynamics of the system is governed by the Liouville type equation $\dot{\rho}=\mc{L}_{\w_t}(\rho)$, where the superoperator $\mc{L}_{\w_t}$ reflects both the unitary dynamics generated by $H$ and the non-unitary contribution induced by the interaction with the  environment.  We further have to assume that the equation for the steady state, $\mc{L}_{\w_t}(\rho)=0$, has a unique solution~\cite{spohn_1978}. The classical Carnot statement is formulated for cycles of  quasistatic (infinitely slow) processes, i.e., successions of stationary states \footnote{Infinitely slow processes are an idealization of processes whose time scales are much longer than the relaxation time \cite{callen}}. In complete analogy to the classical theory quasistatic processes are the only processes considered in the present analysis. 

\section{Thermodynamics of Gibbs equilibrium states}

We begin with a brief review of fundamental thermodynamic concepts for Gibbs equilibrium states,
\begin{equation}
\label{eq01}
\rho=\e{-\beta H}/Z, \quad\text{where}\quad Z=\tr{\e{-\beta H}}\,.
\end{equation}
Here $\beta$ is the inverse temperature of the environment, $\beta=1/T$, and we work in units for which the Boltzmann constant is unity. The thermodynamic entropy is then given by the Gibbs entropy \cite{callen}, $ S=-\tr{\rho\log{\rho}} = \beta \left( E - F \right)$, where $E=\tr{\rho H}$ is the internal energy of the system, and $F=-T\,\log{Z}$ denotes the Helmholtz free energy. For isothermal, quasistatic processes, $\dot{\beta}=0$, the change of thermodynamic entropy $\td S$ becomes
\begin{equation}
\label{eq02}
\td S  = \beta\, \left( \tr{\delta\rho\, H} + ( \tr{\rho\, \delta H } - \td F ) \right) = \beta\, \tr{\delta\rho\, H}\,,
\end{equation}
where $\delta$ denotes an infinitesimal change. Therefore, two forms of energy can be identified \cite{Gemmer2009a}: heat is the change of internal energy associated with a change of entropy; work is the change of internal energy due to the change of an extensive parameter, i.e., change of the Hamiltonian of the system. We have,
\begin{equation}
\label{eq03}
\td E = \delta Q + \delta W \equiv \tr{\delta\rho\, H} + \tr{\rho\, \delta H}\,. 
\end{equation} 
The first law of thermodynamics \eqref{eq03} is a universally valid expression of the conservation of energy. However,
the identification of heat $\delta Q$, and work $\delta W$ \eqref{eq03} is consistent with the second law of thermodynamics for 
quasistatic processes \eqref{eq02} \emph{if}, and as will shortly see, \emph{only if} $\rho$ is a Gibbs state \eqref{eq01}.

It is worth emphasizing that for isothermal, quasistatic processes we have,
\begin{equation}
\label{eq04}
\td S=\beta\, \delta Q \quad \text{and}\quad \td F= \delta W\,,
\end{equation}
for which the first law of thermodynamics takes the form
\begin{equation}
\label{eq05}
\td E=T\,\td S +\td F\,.
\end{equation}
In this particular formulation it becomes apparent that changes of the internal energy $\td E$ can be separated into ``useful" work $\td F$ and an  additional contribution, $T\,\td S$, reflecting the \emph{entropic cost} of the process.

\section{Thermodynamics of non-Gibbsian equilibrium states}

For systems, whose equilibrium states are not described by a Boltzmann-Gibbs distribution \eqref{eq01}, the identification of heat only with changes of the state of the system \eqref{eq03} is no longer possible \cite{Hanggi2008}. Mathematically similar situations have been  studied for classical systems under non-conservative forcing \cite{Oono_1998,Sasa_2001,Seifert2012a}. Such systems relax into so-called nonequilibrium stationary states, and it has been recognized that  not all heat absorbed by the system accounts for the entropic cost \cite{Sasa_2001}. Some contribution to the total heat, coined \emph{housekeeping} energy (or heat) $\delta Q_{hk}$ \cite{Oono_1998,Sasa_2001,Seifert2012a}, fulfills the sole purpose of preventing the system from relaxing into the thermal Gibbs state.  The concept of quantum housekeeping heat has been analyzed carefully in Refs.~\cite{Horowitz_2013,Yuge2013}.

For generic quantum systems, whose thermal equilibrium states are non-Gibbsian, the situation is mathematically analogous.  However, we stress that in the present context we are interested in quantum systems in non-Gibbsian \emph{equilibrium} states, and not in generic nonequilibrium situations. The deviation from the Gibbsian equilibrium is only due to the interaction and correlations between system and environment. In particular, there is no continuous supply of energy from the environment into the system. For instance, it has been seen explicitly in the context of quantum Brownian motion \cite{Horhammer2008} that system and environment are generically entangled.  In such situations the reduced equilibrium state, $\sigma$, of the system only can be written as \cite{Gelin2009},
\begin{equation}
\label{eq_rev_1}
\sigma=\e{-\beta\, (H+\Delta)}/Z_\Delta
\end{equation}
where $H$ is the reduced Hamiltonian and $Z_\Delta$ the modified partition function. A similar situation is encountered in classical systems with non-negligible interacting energy between a system of interest and its environment. For such classical systems it has been shown that $H+\Delta$ can be interpreted as a potential of mean force \cite{Jarzynski2004}, and that the identification of thermodynamic work is subtle \cite{Jarzynski2004}.

Note, however, that generically the physical situation is even more involved in the quantum case. Whereas $\Delta$ for classical systems only includes contributions from interaction energies, solvation energies and classical correlations, for quantum systems $\Delta$ is also governed by quantum correlations. This means that even for situations for which the surface terms such as the interaction energy are vanishingly small, the purely quantum part of $\Delta$ can not necessarily be neglected \cite{Gelin2009}.

Therefore, to formulate thermodynamics consistently the energetic back action due to the correlation of system and environment has  always to be considered carefully \cite{Hanggi2008,Horhammer2008}.  During quasistatic processes parts of the energy exchanged with the environment are not related to a change of the thermodynamic entropy of the system, but rather constitute the energetic price to maintain the non-Gibbsian state, i.e., coherence and correlations between system and environment. 

In complete analogy to stochastic thermodynamics we identify the thermodynamic entropy with the von Neumann entropy \cite{Sasa_2001,tank_2012,Mandal2013a,Horowitz_2013,Sagawa_2015}. With $\sigma$ being the stationary state, we can write
\begin{equation}
\label{eq06}
\begin{split}
 \mc{H} & = -\tr{\s\log{\s}} + \left( \tr{\s\log{\rho}} - \tr{\s\log{\rho}} \right) \\
			     & = \beta ( \mc{E} -(F + T\, \mrm{D}(\s || \rho )) )  = \beta\, ( \mc{E} -\mc{F} ),
	\end{split}
\end{equation}
where $\mc{E}=\tr{\s H}$ is the internal energy of the system, and $\mc{F} \equiv F + T\, \mrm{D}(\s || \rho )$ is the information
free energy~\cite{tank_2012}. Here, $\mrm{D}(\s|| \rho )\equiv \tr{\s\left(\log\s - \log\rho\right)}$ is the quantum relative entropy \cite{Vedral_2002}. Note that it has been shown that $\mc{F}$ is the only thermodynamically consistent definition of a free energy for non-Gibbsian states \cite{tank_2012,Deffner2013}.

As before \eqref{eq02} we now consider isothermal, quasistatic processes, for which the infinitesimal change of the entropy reads
\begin{equation}
\label{eq07}
\begin{split}
		\td\mc{H} & = \beta\, \left[ \tr{\delta\s H} + ( \tr{\s \delta H } - \td\mc{F} ) \right] \\
				  & \equiv \beta\, (\delta Q_\mrm{tot} - \delta Q_{c} ) \,.
\end{split}
\end{equation}
where we identified the total heat as $\delta Q_\mrm{tot}\equiv \tr{\delta\s H} $ and the correlation part as $\delta Q_{c}\equiv\td\mc{F}-\tr{\s \delta H}$. 

The total heat exchanged with the environment has two contributions. The correlation heat is the energetic price that has to be paid to maintain coherence and quantum correlations. The \emph{excess heat} $\delta Q_\mrm{ex}$ is the only contribution that is associated with the entropic cost,
\begin{equation}
\label{eq08}
	\td \mc{H} = \beta\, \delta Q_{\rm ex}, \quad\text{and}\quad \delta Q_{\rm ex} = \delta Q_\mrm{tot} - \delta Q_{c}\,.
\end{equation}
Notice that $\delta Q_\mrm{ex}$ is mathematically analogous to the excess heat for classical systems under non-conservative driving \cite{Sasa_2001}.

Accordingly, the first law of thermodynamics takes the form
\begin{equation}
\label{eq09}
\td \mc{E}=\delta W_\mrm{ex}+\delta Q_\mrm{ex}
\end{equation}
where $\delta W_\mrm{ex}\equiv\delta W+\delta Q_{c}$ is the excess work \cite{tank_2012}, which reduces in the classical limit to the notion analyzed in Ref.~\cite{Jarzynski2004}. Finally, Eq.~\eqref{eq05} generalizes for isothermal, quasistatic processes in generic quantum systems to
\begin{equation}
\label{eq10}
\td \mc{E}= T\,\td\mc{H}+\td\mc{F}\,.
\end{equation}
In the remainder of the present analysis we will show how the universal Carnot statement follows from these generalized thermodynamic relations.

\section{Universal efficiency of quantum Carnot engines}

Imagine a generic quantum system that operates between two heat reservoirs with hot, $T_{\rm hot}$, and cold, $T_{\rm cold}$, temperatures, respectively. Then, the Carnot cycle consists of two isothermal processes during which the systems absorbs/exhausts heat and two thermodynamically adiabatic, i.e., isentropic strokes during which the extensive control parameter $\w$ is varied.


During the first isothermal stroke, the system is put into contact with the hot reservoir. As a result,  the excess heat $Q_{\rm{ex},1}$ is absorbed
at temperature $T_{\rm hot}$ and excess work $ W_{\rm{ex},1}$ is performed,
\begin{equation}
\label{eq11}
	\begin{split}
		W_{\rm{ex},1}  &= \mc{F}(\w_2,T_{\rm hot}) - \mc{F}(\w_1,T_{\rm hot})     \\
		Q_{\rm{ex},1} &= T_{\rm hot}\,(\mc{H}(\w_2,T_{\rm hot}) - \mc{H}(\w_1,T_{\rm hot})).
	\end{split}
\end{equation}
Next, during the isentropic stroke, the system performs work $W_{\rm{ex},2}$ and no excess heat is exchanged with the reservoir, $\Delta\mc{H}=0$. Therefore, the temperature of the engine drops from $T_{\rm hot}$ to $T_{\rm cold}$, 
\begin{equation}
\label{eq12}
	\begin{split}
		W_{ex,2}  & = \Delta\mc{E} = \mc{E}(\w_3,T_{\rm cold}) - \mc{E}(\w_2,T_{\rm hot}) \\
  		 &=\Delta\mc{F} - \left(T_{\rm hot}-T_{\rm cold}\right)\,\mc{H}(\w_3,T_{\rm cold}).
	\end{split}
\end{equation}
In the second line, we employed the thermodynamic identity $\mc{E}=\mc{F} + T\,\mc{H}$, which follows from the definition of $\mc{F}$. During the second isothermal stroke, the excess work 
$  W_{\rm{ex},3} $ is performed on the system by the cold reservoir. This allows for the system to exhaust the excess heat $ Q_{\rm{ex},3} $ at temperature $T_{\rm cold}$. Hence we have
\begin{equation}
\label{eq13}
	\begin{split}
		W_{ex,3} &= \mc{F}(\w_4,T_{\rm cold}) - \mc{F}(\w_3,T_{\rm cold}) 		\\
		Q_{ex,3}  &= T_{\rm cold}(\mc{H}(\w_c,T_{\rm cold}) - \mc{H}(\w_3,T_{\rm cold})).
	\end{split}
\end{equation}
Finally, during the second isentropic stroke, the cold reservoir performs the excess work $ W_{\rm{ex},4} $ on the system. No excess heat is exchanged and the temperature of the engine increases from $T_{\rm cold}$ to $T_{\rm hot}$,
\begin{equation}
\label{eq14}
	\begin{split}
		 W_{ex,4}  & = \Delta\mc{E} = \mc{E}(\w_1,T_{\rm hot}) - \mc{E}(\w_4,T_{\rm cold}) \\
         & = \Delta\mc{F} + \left(T_{\rm hot}-T_{\rm cold}\right)\,\mc{H}(\w_1,T_{\rm hot}).
	\end{split}
\end{equation}
As before, Eq.~\eqref{eq14} reflects the isentropic condition, $\mc{H}(\w_1,T_{\rm hot})=\mc{H}(\w_4,T_{\rm cold})$.

The efficiency of a thermodynamic device is defined as the ratio of  ``output" to ``input". In the present case the ``output'' is the total work performed
during each cycle, i.e., the total excess work, $W_\mrm{ex}= W +  Q_{c}$. There are two physically distinct contributions:  work in the usual sense, $W$, that can be utilized, e.g., to power external devices, and correlation energy, $Q_{c}$, which cannot serve such purposes as it is the thermodynamic cost to maintain the non-Gibbsian equilibrium state. Therefore, the  only thermodynamically consistent definition of the efficiency has to read
\begin{equation}
\label{eq15}
\eta = \frac{\sum_{i}  W_{\rm ex, i} }{ Q_{{\rm ex},1} }  = 1 - \frac{T_{\rm cold}}{T_{\rm hot}}\equiv \eta_\mrm{C},
\end{equation}
which is identical to the classical Carnot efficiency. Earlier analyses did not distinguish between correlation and excess part of the heat, and the efficiency was simply defined as $\eta=W/Q_\mrm{tot}$, see for instance \cite{scully_2003}.

\paragraph*{Example 1}

To illustrate these concepts and to build intuition we now turn to illustrative systems. As a first example we consider quantum Brownian motion, i.e., a harmonic oscillator coupled to an ensemble of harmonic oscillators. In this case the non-Gibbsian equilibrium state of a Brownian particle with mass $m$ becomes \cite{Horhammer2008},
\begin{equation}
\label{new1}
\bra{x}\sigma\ket{y}=\frac{1}{\sqrt{2 \pi\,\la x^2\ra}}\,\e{-\frac{(x+y)^2}{8\la x^2\ra}-\frac{(x-y)^2}{2\hbar^2/\la p^2\ra}}\,,
\end{equation}
where $\la x^2\ra = (D_{pp}+ m \gamma D_{xp})/m^2\gamma^2\omega^2$ and $\la p^2\ra =D_{pp}/\gamma$. Here $\gamma$ is the damping constant, and $D_{pp}$ and $D_{xp}$ are diffusion coefficients, which read in a high temperature expansion $D_{pp}=m \gamma/\beta +m\gamma\beta \hbar^2\,(\omega^2-\gamma^2)/12$ and $D_{xp}=\hbar^2\gamma^2\beta/12$ \cite{Dillenschneider2009a}. To implement the cycle we assume that the angular frequency $\omega$ is controlled externally. 
\begin{figure}
\includegraphics[width=.48\textwidth]{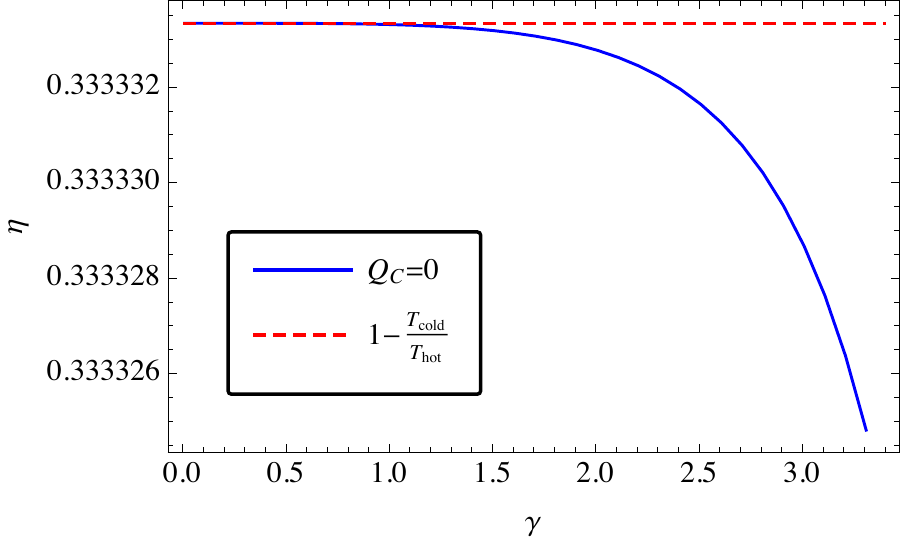}
\caption{\label{fig:eta_harm}(color online) Efficiency of the quantum Carnot cycle for Eq.~\eqref{new1}. The blue, solid line results from $W/Q_\mrm{tot}$, which does not properly account for the correlation energy $ Q_{c}$. The red, dashed line is the Carnot efficiency \eqref{eq15}. Parameters are $T_{\rm cold}=1$, $T_{\rm hot}=1.5$, $\w_1=0.2$, $\w_2=0.6$, $\hbar=1$, and $m=1$.}
\end{figure} 


In Fig.~\ref{fig:eta_harm} we plot the resulting efficiency. We observe that the Carnot efficiency is, indeed, attained for all values of $\gamma$ if one properly accounts for the correlation heat. The blue, solid line is the ratio of work over total heat, $W/Q_\mrm{tot}$. Notice that in this case  $\eta=W/Q_\mrm{tot}$ deviates from $\eta_C$. 

\paragraph{Example 2} 

This deviation becomes even more dramatic in our second example. Consider a quantum particle in a quantum piston as in Fig.~\ref{fig:model}. Such a system can be realized, for instance, as a qubit coupled to an optical cavity with Hamiltonian ($\hbar=1$)
\begin{equation}
\label{eq16}
H = \frac{\w_q}{2}\,\sig{z} + \w_b a^{\dg}a + g\,\sig{x}\otimes ( a^{\dg} + a )\,.
\end{equation} 
Here, $a$, $a^{\dg}$ are the annihilation and creation operators of bosonic modes with frequency 
$\w_b$ \cite{kumar_2013}. The base frequency $\w_b$ is the  parameter to be changed, which can be experimentally realized by varying the laser in the cavity. Pauli matrices $\sig{z}$, $\sig{z}$ represent a two level atom with energy $\w_q/2$
and its coupling to the cavity \cite{Zhang_2014}. Finally, the last term $\sig{x}\otimes(a^{\dg}+a)\equiv \sig{x} \otimes \hat{x}$ describes the interaction of qubit and piston and can be interpreted as an intra-system  non-conservative forcing of strength $g$.
\begin{figure}
\includegraphics[width=.52\textwidth]{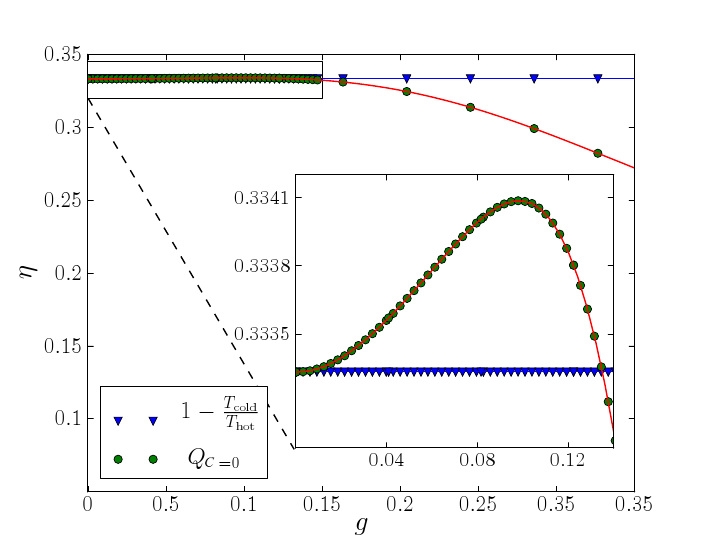}
\caption{\label{fig:eta}(color online) Efficiency of the quantum Carnot cycle for Eq.~\eqref{eq16} with Eqs.~\eqref{eq17} and \eqref{eq18}. The blue triangles are a numerical verification of \eqref{eq15}, whereas the green circles result from $W/Q_\mrm{tot}$, which does not properly account for the correlation energy $ Q_{c}$. The inset is a magnification for small values of $g$. Parameters are $\w_q=1$, $\gamma_q=\gamma_b=0.05$, $T_{\rm cold}=1$, $T_{\rm hot}=1.5$, $\w_1=0.2$, $\w_2=0.6$. 
}
\end{figure} 

Finally, the thermal reservoirs are phenomenologically modeled by a Lindblad master equation~\cite{breuer_book}, 
$\mc{L}(\rho) = -i\left[ H,\rho \right] + \mc{D}_q \left( \rho \right) + \mc{D}_b \left( \rho \right)$, where
\begin{equation}
\label{eq17}
	\begin{split}
        \mc{D}_q \left( \rho \right) &= \gamma_q\, (N_{\w_q}+1)\left(\s_{-}\rho\s_{+}-\frac{1}{2}\{\s_{-}\s_{+},\rho\} \right) \\
        &+ \gamma_q\, N_{\w_q}\left(\s_{+}\rho\s_{-}-\frac{1}{2}\{\s_{+}\s_{-},\rho\} \right),
	\end{split}
\end{equation}
with ladder operators for the atom, $\s_{\pm}$, and
\begin{equation}
\label{eq18}
	\begin{split}
        \mc{D}_b \left( \rho \right) &= \gamma_b\, (N_{\w_b}+1)\left(a_{-}\rho a^{\dg}-\frac{1}{2}\{aa^{\dg},\rho\} \right) \\
        &+ \gamma_b\, N_{\w_b}\left(a^{\dg}\rho a-\frac{1}{2}\{a^{\dg}a,\rho\} \right).
	\end{split}
\end{equation}
Here, $N_{x}=1/(\e{\beta x}-1)$ and $\gamma_q$, $\gamma_b$ are fermionic and bosonic coupling constants, respectively. Lindblad master equations are generally applicable only to describe quantum system weakly coupled to the (classical) environment. For the present case this assumption is justified as the Hamiltonian \eqref{eq16} describes a generic quantum optomechanical system, for which Lindblad master equations have been proven to be adequate~\cite{breuer_book,Carmichael1991}. Moreover, we do not have to account for dynamical corrections as we are only interested in quasistatic, i.e., infinitely slow processes. It is worth emphasizing that from microscopic treatment one would expect that the interaction between the two subsystems changes the individual dissipators~\cite{Huelga}. However, for a macroscopic Lindblad master equation those corrections would force the system to relax into a Gibbs state. For the present purposes, we have specifically chosen a phenomenological system which does not relax into a Gibbs equilibrium state.

In  this model the two subsystems are coupled to the thermal reservoir independently. However, they ``feel" each other through the direct interaction. Only in the limit $g \to 0$ the steady state is a Gibbs state~\cite{breuer_book}.  For finite interaction qubit and cavity are correlated and they share information. The thermodynamic price for maintaining  this correlation during the thermodynamic cycle is the correlation energy $ Q_{c}$~\eqref{eq07}.

Figure~\ref{fig:eta} plots the resulting efficiency \eqref{eq15}. We observe again that the classical Carnot efficiency is, indeed, attained for all values of $g$. The green circles are the ratios of work over total heat, $W/Q_\mrm{tot}$. Notice that in this case the Carnot statement appears to be violated as $\eta=W/Q_\mrm{tot}$ can be larger or smaller than $\eta_\mrm{C}$ as a function of $g$. This apparent violation is not physical, but is rather rooted in an thermodynamically inconsistent identification of the excess heat.

\vspace*{1em}

\section{Concluding remarks}
The present studied analyzed the thermodynamics of non-Gibbsian quantum heat engines -- devices that operate cyclically in 
non--Gbbsian equilibrium states. We investigated the thermodynamic processes underlining such nanodevices
and concluded that it is impossible to harness quantum correlations in quasistatic processes to enhance the maximum efficiency 
of such devices. Instead one has to modify the definition of heat, and account for the correlation energy necessary to maintain coherence and correlations. In conclusion, we showed that the Carnot statement of the second law is universally valid also for quantum heat engines.

\acknowledgments
It is a pleasure to thank Wojciech H. \.Z{}urek, Christopher Jarzynski, and Dibyendu Mandal for stimulating discussions. We gratefully acknowledge Marta Paczy\'n{}ska, who put our theoretical ideas into a artistically pleasing form for Fig.~\ref{fig:model}. This work was supported by the Polish Ministry of Science and Higher Education under project Mobility Plus, $1060/$MOB$/2013/0$, 
(BG). SD acknowledges financial support from the U.S. Department of Energy through a LANL Director's Funded Fellowship.

\bibliography{carnot}

\end{document}